# Towards blind user's indoor navigation: a comparative study of beacons and decawave for indoor accurate location

Prabin Sharma[1,3] Sambad Bidari[1] Kisan Thapa[4] Antonio Valente[2] Hugo Paredes[1,2]

[1] Universidade de Tras-os-Montes e Alto Douro, Vila Real, Portugal

[2] INESC TEC, Porto, Portugal

[3] Kathmandu University, Dhulikhel, Nepal

[4] Pokhara University, Pokhara, Nepal

*Abstract*— there are many systems for indoor navigation specially built for visually impaired people but only some has good accuracy for navigation. While there are solutions like global navigation satellite systems for the localization outdoors, problems arise in urban scenarios and indoors due to insufficient or failed signal reception. To build a support system for navigation for visually impaired people, in this paper we present a comparison of indoor localization and navigation system, which performs continuous and real-time processing using commercially available systems (Beacons and Decawave) under the same experimental condition for the performance analysis. Error is calculated and analyzed using Euclidean distance and standard deviation for both the cases. We used Navigine Platform for this navigation system which allows both Tri-lateration as well as Fingerprinting algorithms. For calculating location we have used the concept of Time of Arrival and time of difference of arrivals. Taking into concern about the blind people, location is important as well as accuracy is necessity because small measurement in the walk is important to them. With this concern, in this paper, we are showing the comparative study of beacons and Decawave. The study and the accuracy tests of those systems for the blind people/user's in navigating indoor are presented in this paper.

## I. INTRODUCTION

In a pervasive computing world, location information is important. For example, location information can be used to help users find what they need and where it is from the current location of the users. A tracking system can be used to guide the navigation in the urban environment indoor area [1]. Similarly, a navigation system is used to guide users to a certain location. For example, a car navigator is used to guide a driver to a destination based on the current location of the vehicle in real-time or turn-by-turn. The location given to the navigator is typically calculated by Global Position System (GPS) receiver that receives reference radio signals from GPS satellites. Because of GPS system the outdoor navigation is so much successful but if we compare with indoor navigation, it's not as successful as outdoor navigation due to signal attenuation by construction material of buildings [2]. People can utilize an indoor navigation system to locate devices throughout a building, tourists can use it as a tour guide in a museum, or fire fighters can use it to find an emergency exit in the smoky environments where it is difficult to see the way. A navigation system and location information is essential for the visually impaired people, to locate themselves and navigate them to the desired location.

Navigation for a visually impaired people is totally a big challenge. There are many solutions for blind navigation that have arisen for the last 30years. Some are now widely used as NavCog from CMU , a navigational cognitive assistance for visually impaired people [3] which provides turn by turn assistance to the visually impaired person according to the real time localization in a large space. Microsoft also developed the solution for localizing for visually impaired people guiding through audio by the Application named Soundscape which helps the user to enrich ambient awareness. There are some system like ISEABlind [4] which was developed at INESTEC and UTAD uses artificial vision, assisted navigation, detailed spatial perception and different kinds of sensors to give audio feedback to the user for the smooth mobility of visually impaired user.

In developing technology for indoor localization, we have recently begun exploring commercially available state of the art localization technologies. There are different systems used in the Indoor navigation system. As mentioned by [5] there are technologies like Infrared(IR), Ultrasound, Radio frequency Identification (RFID), Wireless Local Area Network (WLAN), Bluetooth, Sensor networks, Ultra-Wideband (UWB) ,Magnetic signals . Out of them only few of the technology are commercially available and which are available are costly and hard to install and deploy.

The technology of beacons are low cost and high spatial selectivity however the accuracy is the major factor because typical walking speed are likely to change which effects the localization of the place or object[6]. Decawave has taken over the beacons in terms of accuracy as well as low-cost real-time location system (RLTS) applications. The aim of this paper is to compare the accuracy of indoor navigation system using beacons and decawave which can aid in the navigation of a visually impaired user for indoor purpose.

While there are solutions like global navigation satellite systems (GNSS) for the localization outdoors, problems arise in urban scenarios and indoors due to insufficient or failed signal reception. For indoor use, multiple alternative localization concepts exist that are suited for different use cases [7]. Through the years there has been a shift in their preferences as most of the blind/visually impaired people wish to be a bit more independent. People are shifting towards the next big thing for mobility, that being assistive technology. Emerging smart services like RFID, WLAN, Beacon, Decawave etc. are being used to locate and navigate the visually impaired people in the indoor environment which differ from a lack of cost-effective and deployment. To address this concern, in this paper we present a comparison of indoor localization and navigation system, which performs continuous and real-time processing using beacons and decawave.

## II. BACKGROUND

The development of outdoor navigation system has been so rapid as compared to indoor navigation system. Accuracy in localization is very important for visually impaired people, even the small error can deviate the person to undesired place. Comparing the accuracy test for indoor navigation system plays vital role to ease the visually impaired people to localize.

An experiment was performed using beacons for indoor navigation system to analyze the positioning accuracy corresponding to number of beacons in Daejeon, South Korea [8] . Experiment was done in the space of 100m*100m in which 10-100 beacons were deployed in grid and random topology. They found out the more accurate result when there is more beacons.

Another research was conducted in one floor of the building of the Faculty of Informatics and Management, University of Hradec Kralove where beacons were deployed [9] . They combined Wi-Fi network and beacons together in indoor navigation system to improve the accuracy of localization. The outcome show that the accuracy can be improved by 23% with less variance.

In order to find the accuracy of UWB positioning systems various research work have been conducted. A research in Lopsi lab was conducted to compare the accuracy of Decawave as well as Bespoon location system [10] . As a result they found out the standard deviation of Bespoon and Decawave system were 11cm and 5.5 cm respectively which signifies Decawave system is more accurate than Bespoon System.

## III. EXPERIMENTAL SETUP

The Experiment was performed in the Human Computer Interaction Lab of University of Trs-os-Montes and Alto Douro, Vila real, Portugal. Both systems: Beacons and Decawave Dw1001 were implemented by deploying the modules in the same position one after another. The area of Lab is 10m*8m. The main goal of this paper is to compare the indoor navigation system using Beacons technology and Decawave modules. Decawave module and Beacons were deployed on the wall of the lab which was quite far from the metallic ceiling which could have attenuated the signal given by Transmitter.

### A. Beacons

*1) Hardware used:* In this system we have two parts. They are as Transmitter and Receiver.

  *a) Transmitter Unit:* As a transmitter we used IBKS150 which will transmit signal to the receiver. It uses Nordic Semiconductors nrf51822 chipset which is a powerful and also highly flexible multiprotocol SoC ideally suited for Bluetooth Low Energy. It uses CR2477 coin cell battery which is small in size and long lasting.

  *b) Receiver:* As a receiving unit we used Samsung galaxy S5 android 6.0 which has Bluetooth 4.0 (BLE).The mobile phone receives signals strength from the transmitter (Beacons) and use trilateration to find the location of user.

Beacons are small radio Bluetooth transmitters. It's kind of like a lighthouse: it repeatedly transmits a single signal that other devices can see. Bluetooth low Energy (BLE) Beacon is a small device which can be attached in any surface which can communicate to nearby devices using Bluetooth.

There are various positioning techniques that have been proposed till date. Among them most common position techniques are Trilateration, Triangulation and Fingerprinting [11] .

*2) Algorithm Used:* We used Navigine Platform for this navigation system which allows both Trilateration as well as fingerprinting algorithms. Fingerprinting algorithm allow to use both Wi-Fi infrastructure and Beacons. Due to attenuation
, Trilateration requires a correction factor whereas Fingerprinting algorithm considers the correction factor already in the database creation process [12] .

  *a) Fingerprinting:* Fingerprinting is the location positioning technique that first collects the fingerprints (features) from the scene and the location of the object is found out by matching the online measurement with the one from the location fingerprints from the database [13]. In order to find the device location, the device collects the samples of signals of the surrounding and the device calculates RSSI values. The RSSI samples are compared with the previously recorded RSSI samples and the one which is best match with the unique pattern of sample is taken in order to find the device location [14].

In fingerprinting process there are Two Phases: Online phase and Offline phase.

- Offline Phase: In the Offline phase, the data of the location coordinate along with the respective signal strength is collected by going in the site on foot in the radio environment.
- Online Phase: In the online phase, the currently data collected is compared with the samples which are already collected to find the estimate location.

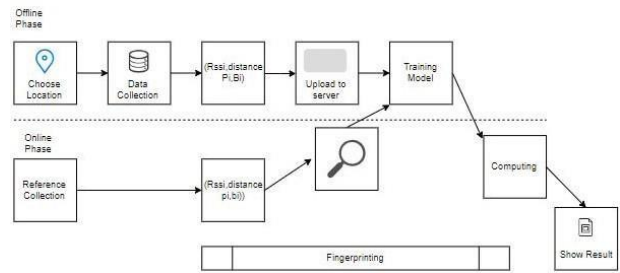

Fig. 1. Principle of Fingerprinting Algorithm

## B. Decawave

*1) Hardware used:* In this system we have two parts. They are as Transmitter and Receiver.

*a) Transmitter Unit:* In this Decawave System we used Tags as the transmitting unit which is used to send signal to the receiver.

*b) Receiver:* We use Anchor as a receiver which gains signal from the tags which are located at the fixed location indoor. It measures the time between the transmission and the reception of the signals.

*2) Location Calculation:* There are two operation modes in this system: Time of Arrival (ToA) and Time Difference of Arrival (TDoA).

*a) Time of Arrival (ToA):* ToA can be used to find the location with 3 transmitters (or more for better accuracy, 4 for 3D). In this mode, Tag periodically performs the sequence of two way communication range. At the end of those sequence, all the measured distances between tag and each of the anchors is sent to the server, where this information is forwarded to the Location Engine. Location Engine (LE) provides an estimate of tag's position, coordinates and their accuracy.[15]

*b) Time Difference of Arrival (TDoA):*
In this operating mode, the received signals are cross-correlated to determine the TDOA and a set of nonlinear equations are solved to produce the location estimate. It is somewhat more versatile than ToA[16] .This method does not require the time that the signal was sent from the target, only the time the signal was received and the speed that the signal travels. Once the signal is received at two reference points, the difference in arrival time can be used to calculate the difference in distances between the target and the two reference points.[17]

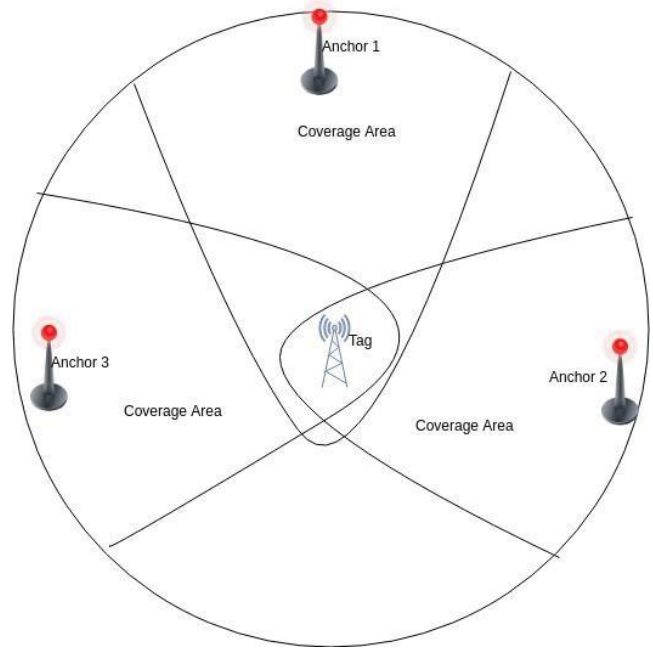

Fig. 3. Principle of Time Difference of Arrival Algorithm

In fig.2 and Fig.3 we can see that how the algorithm of Time of Arrival (TOA) and Time difference of Arrival works (TDOA).

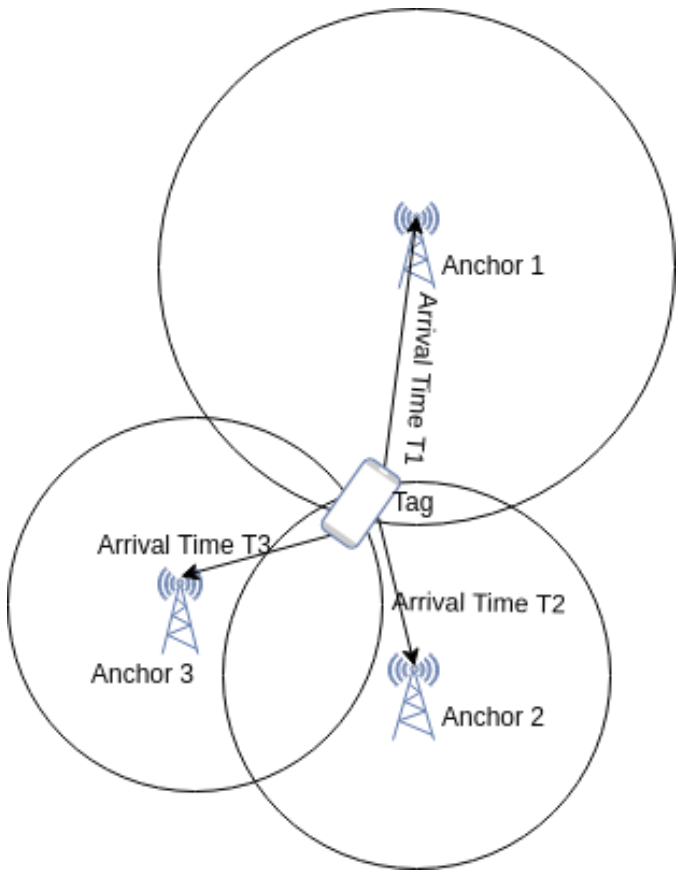

Fig. 2. Principle of Time of Arrival Algorithm

## IV. TEST MEASUREMENT

For the beacons we used Navigine Platform to find the location of a device and For Decawave DW1001 module we used Decawave Application to configure anchor and tags hence to find the user location. Figure 4 shows the floor plan in which blue (X) represents the position where infrastructure is deployed. A, B, C, D and E are the Position which is taken for accuracy test.

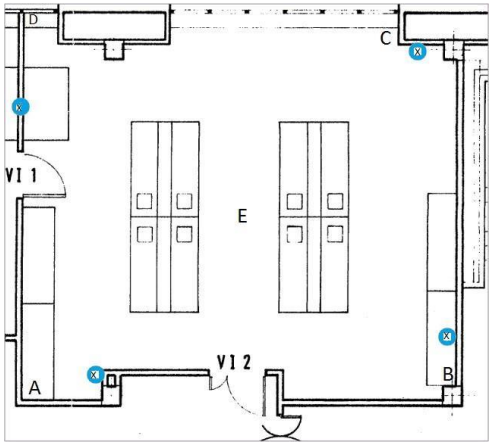

Fig. 4. Floor plan of HCI lab with infrastructure deployed

### A. Accuracy Test of Beacon system

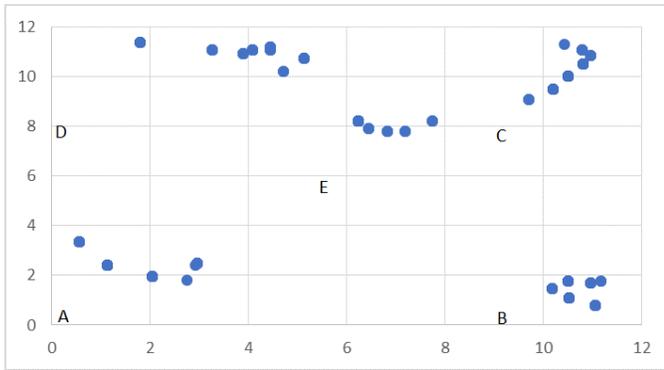

Fig. 5. Calibration of location using Beacon Navigation Technology

*1) Error Calculation for Beacons Technology using Euclidean Distance:* We used Euclidean Distance measurement to find the distance error. The distance error is calculated at points A, B, C, D and E where data are collected. The distance formula for Euclidean is described as:

Euclidean distance = $\sqrt{(x_2 - x_1)^2 + (y_2 - y_1)^2}$

Where (x1, y1) is accurate data (reference point) and (x2, y2) is mean of data from the cluster of data obtained.

The Euclidean distance between the mean of cluster of points and actual point is Distance error which is obtained as below:

- Error obtained for point A (0,0)= 2.9m
- Error obtained for point B(9.20,0)=2.82m
- Error obtained for point C(9,7.85)=3.0m
- Error obtained for point D(0,7.85)=3.5m
- Error obtained for point E(5,5.5)=1.5m
- Average Error= 2.7m

### B. Accuracy Test of Decawave module Navigation System

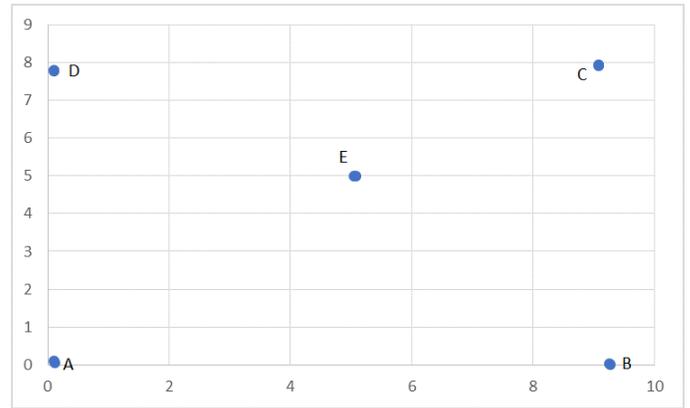

Fig. 6. Calibration of location using Decawave Technology

*1) Error Calculation:*

- Error obtained for point A (0,0)= 13cm
- Error obtained for point B(9.20,0)=8cm
- Error obtained for point C(9.20,7.85)=10cm
- Error obtained for point D(0,7.85)=12cm
- Error obtained for point E(5,5.5)=5cm
- Average Error= 9.6cm

### C. Analysis using Standard Deviation

Standard Deviation plays an important role in statistical analysis. As standard deviation is a measure to find the deviation or dispersion from the mean value [18]. so we calculated the standard deviation of X and Y coordinates of the Actual points (A,B,C,D and E) for Beacons Technology and Decawave Technology.
Fig.7 and Fig8 shows the deviation of X and Y coordinates with the standard deviation on the Y axis with the unit meter(m). Clearly from the figure we can say that the standard deviation in the Lab space using Beacon technology is higher than the standard deviation obtained using Decawave Technology.

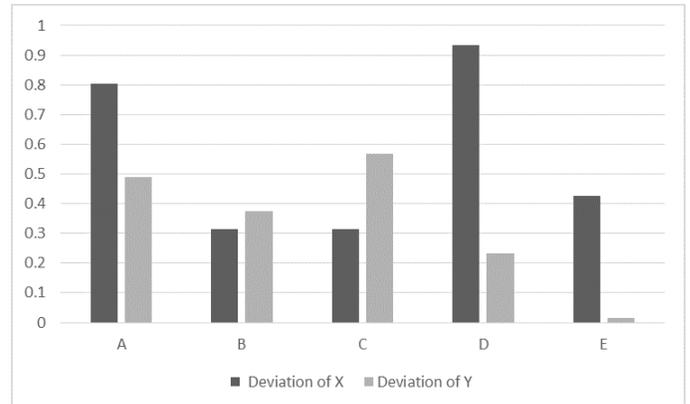

Fig. 7. Standard Deviations in X and Y coordinates while using Beacon Technology

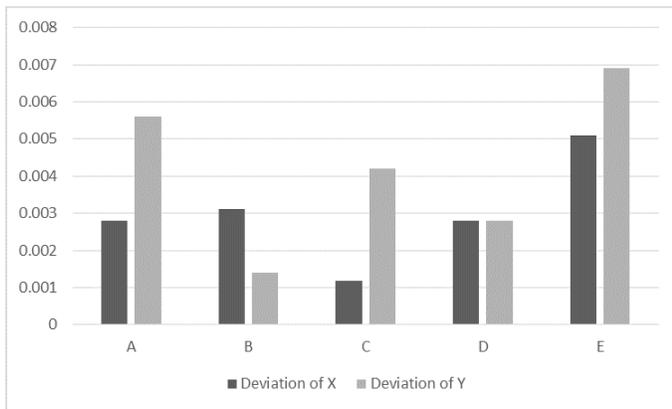

Fig. 8. Standard Deviations in X and Y coordinates while using Decawave Technology

The data are more spread in Beacon technology which results in more deviation with respect to mean. More deviation results in more error which will be a challenge to find the actual location of the user. Whereas in case of Decawave Technology the deviation obtained is very less compared to Beacon Technology which is really good for Blind people to navigate with support system.

## V. RESULT AND DISCUSSION

We tested Beacon Technology as well as Decawave Technology for the accuracy test in-order to make support system for the blind Navigation system. We used Standard Deviation and Euclidean distance to compare both system. We found out Decawave system is really good system with minimal error because of its Algorithm implemented in the system. Visually impaired people needs navigation system with minimal error so that they can reach to their target destination without any obstacles. Any significant error in navigation can lead the visually impaired people to reach to false destination which will decrease the efficiency as well may harm the user. So Decawave Technology is chosen over Beacon Technology for the construction of Blind navigation support system.

## VI. FUTURE WORK

Although we only tested the system accuracy test only but our aim is to make Blind navigation support system mobile application using the system which gives the best accuracy. In the future we can make a User friendly mobile application for visually impaired people using Decawave Technology in order to help them to navigate them in indoor locations. We will try to make an automatic speech helper for user to make them aware where are they are located which helps them to navigate using voice command. They will also be able to command through voice where they desire to go inside the indoor location.

## REFERENCES

[1] Ehud Mendelson. System and method for providing indoor navigation and special local base service application for malls stores shopping centers and buildings utilize rf beacons, October 21 2014. US Patent 8,866,673.
[2] Anja Bekkelien, Michel Deriaz, and Stéphane Marchand-Maillet. Bluetooth indoor positioning. *Master's thesis, University of Geneva*, 2012.
[3] Dragan Ahmetovic, Cole Gleason, Chengxiong Ruan, Kris Kitani, Hironobu Takagi, and Chieko Asakawa. Navcog: a navigational cognitive assistant for the blind. In *Proceedings of the 18th International Conference on Human-Computer Interaction with Mobile Devices and Services*, pages 90–99. ACM, 2016.
[4] Tânia Rocha, Hugo Fernandes, Arsénio Reis, Hugo Paredes, and João Barroso. Assistive platforms for the visual impaired: bridging the gap with the general public. In *World Conference on Information Systems and Technologies*, pages 602–608. Springer, 2017.
[5] Yanying Gu, Anthony Lo, and Ignas Niemegeers. A survey of indoor positioning systems for wireless personal networks. *IEEE Communications surveys & tutorials*, 11(1):13–32, 2009.
[6] Sudarshan S Chawathe. Low-latency indoor localization using bluetooth beacons. In *Intelligent Transportation Systems, 2009. ITSC'09. 12th International IEEE Conference on*, pages 1–7. IEEE, 2009.
[7] Ramsey Faragher and Robert Harle. Location fingerprinting with bluetooth low energy beacons. *IEEE journal on Selected Areas in Communications*, 33(11):2418–2428, 2015.
[8] Myungin Ji, Jooyoung Kim, Juil Jeon, and Youngsu Cho. Analysis of positioning accuracy corresponding to the number of ble beacons in indoor positioning system. In *Advanced Communication Technology (ICACT), 2015 17th International Conference on*, pages 92–95. IEEE, 2015.
[9] Pavel Kriz, Filip Maly, and Tomas Kozel. Improving indoor localization using bluetooth low energy beacons. *Mobile Information Systems*, 2016, 2016.
[10] Antonio Ramón Jiménez and Fernando Seco. Comparing decawave and bespoon uwb location systems: Indoor/outdoor performance analysis. In *IPIN*, pages 1–8, 2016.
[11] Christoph Fuchs, Nils Aschenbruck, Peter Martini, and Monika Wieneke. Indoor tracking for mission critical scenarios: A survey. *Pervasive and Mobile Computing*, 7(1):1–15, 2011.
[12] Ting Wei and Scott Bell. Indoor localization method comparison: Fingerprinting and trilateration algorithm. *University of Saskatchewan. Accessed March*, 24:2015, 2011.
[13] Hui Liu, Houshang Darabi, Pat Banerjee, and Jing Liu. Survey of wireless indoor positioning techniques and systems. *IEEE Transactions on Systems, Man, and Cybernetics, Part C (Applications and Reviews)*, 37(6):1067–1080, 2007.
[14] Meng Wang. *Indoor Navigation Systems Based On iBeacon Fingerprinting*. PhD thesis, Vanderbilt University, 2015.
[15] Oleg Tabarovsky, Vladimir Maximov, Evgeny Emelyanov, Dmitry Matitsin, Anatoly Zimin, Evgeniya Kalnina, Alexander Avdeev, and Irina Novikova. Hihg-accuracy indoor positioning system with decawave transievers and auto-calibration.
[16] KC Ho and YT Chan. Solution and performance analysis of geolocation by tdoa. *IEEE Transactions on Aerospace and Electronic Systems*, 29(4):1311–1322, 1993.
[17] Li Cong and Weihua Zhuang. Hybrid tdoa/aoa mobile user location for wideband cdma cellular systems. *IEEE Transactions on Wireless Communications*, 1(3):439–447, 2002.
[18] Mohini P Barde and Prajakt J Barde. What to use to express the variability of data: Standard deviation or standard error of mean? *Perspectives in clinical research*, 3(3):113, 2012.